\documentclass[prl,twocolumn,tightenlines,showpacs,nofootinbib]{revtex4}
\usepackage{bm,dcolumn,amsmath,graphicx,amsfonts,amssymb}
\usepackage{epsfig}

\begin{document}
\title{Revisiting parity non-conservation in cesium}
 
\author{V. A. Dzuba, J. C. Berengut, V. V. Flambaum, and B. Roberts}
\affiliation{School of Physics, University of New South Wales,
Sydney, NSW 2052, Australia}

\date{ \today }

\begin{abstract}
We apply the sum-over-states approach to calculate partial contributions to
the parity non-conservation (PNC) in cesium [Porsev {\em et al},
Phys. Rev. D {\bf 82}, 036008 (2010)]. We have found significant
corrections to two non-dominating terms coming from the contribution of
the core and highly excited states ($n>9$, the so called {\em tail}).
When these differences are taken into account the result of Porsev {\em
  et al}, $E_{\rm PNC} = 0.8906\,(24) \times 10^{-11}i(-Q_W/N)$ changes to
$0.8977\,(40)$, % \times 10^{-11}i(-Q_W/N)$  
coming into good agreement with our previous
calculations, $0.8980\,(45)$. The interpretation of the PNC measurements in cesium
still indicates reasonable agreement with the standard model
($1.5\,\sigma$), however gives new constraints on physics beyond it.
\end{abstract}
\maketitle

\section{Introduction}

The search for new physics beyond standard model using parity
non-conservation (PNC) in atoms culminated in 1997 when Boulder group
reported very accurate measurements of the PNC $6s-7s$ amplitude in
cesium~\cite{Wood}. The experimental uncertainty was only
0.35\%. Interpretation of the measurements based on early
calculations by Novosibirsk~\cite{Nov} and Notre Dame~\cite{ND} groups
indicated good agreement with the standard model. However, the
declared theoretical accuracy of these early calculations (1\%) didn't
match the accuracy of the measurements. 
Bennett and Wieman re-analyzed the accuracy of the calculations by
comparing calculated observables with new experimental
data~\cite{BW}. They pointed out that many discrepancies between theory
and experiment were resolved in favor of theory and suggested that
actual accuracy of the calculations~\cite{Nov,ND} was 0.4\%. This
lead to about 2.3\,$\sigma$ deviation of the value of the weak charge of
the cesium nucleus from the prediction of the standard model. The
discrepancy were resolved when Breit~\cite{DeBreit,DzBreit} and
radiative (see, e.g~\cite{QED} and references therein) corrections
were included into calculations. On the other hand, new calculations
of the correlations~\cite{KPT01,DFG02} didn't change the old results
of \cite{Nov,ND} and rather confirmed the suggestion made in \cite{BW}
that their accuracy was high. The new value of the weak nuclear charge
was about 1$\,\sigma$ smaller than the prediction of the standard model
which should be considered as good agreement.

The situation changed when the latest, most sophisticated calculations
of the PNC in Cs were reported by Porsev {\em et
  al}~\cite{Porsev}. The authors of this work used the sum-over-states
approach and applied the coupled-cluster with single, double and
valence triple excitations for the leading terms in the sum. They
claimed 0.27\% uncertainty of the calculations while their
correlated PNC amplitude was about 0.9\% smaller than in previous
calculations. This led to perfect agreement with the standard model,
with central points for the weak nuclear charge extracted from the
measurements and predicted by the standard model coinciding exactly:
$Q_W(^{133}{\rm Cs}) = -73.16(29)_{\rm exp}(20)_{\rm theor}$ and
$Q_W^{\rm SM}(^{133}{\rm Cs}) = -73.16(3)$~\cite{Porsev}. 
The smaller value of the correlated PNC amplitude is attributed in
\cite{Porsev} to the role of higher-order correlations.

Although all old and new calculations of the PNC in Cs lead to
agreement with the standard model, the results of \cite{Porsev} have
important implications imposing strong constraints on new physics
beyond the standard model. Therefore, it is worth studying further the
reasons for the difference in the calculations. 

The authors of
\cite{Porsev} paid great deal of attention to the leading terms,
performing very sophisticated calculations for them and demonstrating
high accuracy by comparing with available experimental data. 
The uncertainty for the minor terms was assumed to be 10\% based on the
spread of the values in different approximations.
The sum-over-states approach used in \cite{Porsev} has an important
shortcoming. Calculation of each term in the sum is practically
independent of others and therefore the high accuracy for the leading
terms does not guarantee high accuracy for the sum. 

In this paper we use the sum-over-states approach to study possible
reasons for the difference between the results of \cite{Porsev} and
previous calculations~\cite{Nov,DFG02}. We assume that the main
term was calculated correctly in \cite{Porsev} and focus our attention
on the minor terms such us contribution of the core states and highly
excited (tail) states. We include core polarization and correlation
corrections into the core and tail contributions and find significant
difference for both these terms between our calculations and those
reported in~\cite{Porsev}. Our core contribution has different sign
while being similar in value. 
We have agreement with \cite{Porsev} for the tail contribution when
only core polarization effects are taken into account. However,
further inclusion of Bruekner-type correlations increase the PNC
amplitude beyond the 10\% uncertainty for the tail assumed in \cite{Porsev}.

If core and tail contributions of \cite{Porsev}
are replaced by the findings of present work the resulting PNC
amplitude comes into excellent agreement with previous calculations. The
application of our calculations to the analysis of the PNC
measurements in Cs leads to a value of weak nuclear charge that is about
1.5\,$\sigma$ smaller than the value predicted by the standard model.
While the PNC amplitude found in this work is practically the same as
in \cite{DFG02,QED}, the apparent increase in deviation from the standard
model (1\,$\sigma$ in \cite{DFG02,QED} when proper values of the Breit,
radiative and neutron skin corrections are added) is due to smaller
uncertainty. This smaller uncertainty is mostly due to small
uncertainty of the main term which we have taken from
Ref.~\cite{Porsev} without re-analysis.

The PNC amplitude calculated in this work gives new constraints on
physics beyond the standard model.

\section{Calculations}

The PNC amplitude of an electric dipole transition between the $6s$
and $7s$ states of cesium can be written as
\begin{eqnarray}
   E_{\rm PNC}  &=&  \sum_{n} \left[
\frac{\langle 6s |  H_{\rm PNC}| np_{1/2}  \rangle
      \langle np_{1/2} | {\bm d} | 7s \rangle}{E_{6s} - E_{np_{1/2}}}\right.
\nonumber \\
      &+&
\left.\frac{\langle 6s | {\bm d} | np_{1/2}  \rangle
      \langle np_{1/2} | H_{\rm PNC} | 7s \rangle}{E_{7s} - E_{np_{1/2}}} \right],
\label{eq:pnc}
\end{eqnarray}
where ${\bm d} = -e\sum_i {\bm r_i}$ is the electric dipole operator,
$H_{\rm PNC}$ is the operator of a P-odd CP-even weak interaction.
\begin{eqnarray}
     H_{\rm PNC} &=& -\frac{G_F}{2\sqrt{2}}  
   Q_W \gamma_5\rho({\bm r}) \ ,
\label{e1}
\end{eqnarray}
$G_F \approx 2.2225 \times 10^{-14}$ a.u. is the Fermi constant of
the weak interaction, and $Q_W$ is the nuclear weak charge.

Expression (\ref{eq:pnc}) is exact if states $6s$, $7s$, $np_{1/2}$ label
many-electron physical states of the atom. Then $6s$ is the ground
state and summation goes over excited states of the opposite parity
and the same total angular momentum $J=1/2$. In practical calculations
equation (\ref{eq:pnc}) is reduced to one with single-electron
orbitals and single-electron matrix elements. It looks very similar to
(\ref{eq:pnc}) but with a few important differences. (a) All states
($6s$, $7s$ $np_{1/2}$) are now single-electron states obtained
with the use of the Hartree-Fock method. (b) Many-electron effects are
reduced to redefinition of the single-electron orbitals and
interaction Hamiltonians. For example, the inclusion of the core
polarization effect leads to redefinition of the interaction
Hamiltonian. For the weak interaction we have $H^{\prime}_{\rm PNC} =
H_{\rm PNC} + \delta V_{\rm PNC}$, where $\delta V_{\rm PNC}$ is the
correction to the self-consistent potential of the atomic core due to
the effect of weak interaction $H_{\rm PNC}$. For the electric dipole
interaction we have similar expression ${\bm d}^{\prime} = {\bm d} +
\delta V_d$. (c) Summation in (\ref{eq:pnc}) now goes over the complete
set of single-electron states including states in the core. Extending
summation to the core states corresponds to inclusion of highly excited
autoionizing states. (d) The expression (\ref{eq:pnc}) via
single-electron states is approximate. Its accuracy is determined by
how the many-body effects are included.

Equation (\ref{eq:pnc}) implies the sum-over-states approach which we
are going to study in this paper. As mentioned above, high
accuracy for the leading terms does not guarantee high accuracy
for the sum. 
To test the total sum we use an alternative approach which we have
used in our previous PNC calculations~\cite{Nov,DFG02}. This approach is
based on the solving of differential equations. 

The PNC amplitude (\ref{eq:pnc}) can be rewritten as
\begin{equation}
    E_{\rm PNC} = \langle \delta\psi_{6s}|{\bm d}|\psi_{7s}\rangle +
\langle \psi_{6s}|{\bm d}|\delta\psi_{7s}\rangle,
\label{eq:pnc2}
\end{equation}
where the $\psi$ and $\delta\psi$ are single-electron orbitals and
$\delta\psi_a$ is the correction to the wave function $\psi_a$ due to 
the weak interaction 
\begin{equation}
\delta\psi_a = \sum_{n} 
\frac{\langle a | H^{\prime}_{\rm PNC} | np_{1/2}  \rangle}{E_{a} -
  E_{np_{1/2}}}\langle np_{1/2} | .
\label{eq:dpsi}
\end{equation}
It is easy to see that this correction to the wave function satisfies
the differential equation
\begin{equation}
(\hat H_0 - E_a)\delta\psi_a = - H^{\prime}_{\rm PNC}\psi_a.
\label{eq:edpsi}
\end{equation}
Here $\psi_a$ is the eigenstate of the $\hat H_0$ Hamiltonian, which
is in our case the relativistic Hartree-Fock (RHF) Hamiltonian. The
equations (\ref{eq:edpsi}) have a form of the RHF equations with the
right-hand side. Solving differential equation (\ref{eq:edpsi}) for the
$6s$ and $7s$ states of cesium and using (\ref{eq:pnc2}) to calculate
the PNC amplitude does not require a complete set of single-electron
states. It is usually numerically more accurate than the use of the
sum-over-states approach. In present work we use it as an
independent test of the calculations.

To perform the summation in (\ref{eq:pnc}) we use the B-spline basis
set first presented in Ref.~\cite{B-splines}. We use 100 B-splines in
each partial wave in the cavity of radius 75 $a_B$. The cavity radius
is taken to be the same as in Ref.~\cite{Porsev}. Its value is
dictated by the need to have the dominating states be as close to
physical (spectroscopic) states as 
possible. The most important intermediate states, according to
\cite{Porsev}, are the $6p_{1/2}$, $7p_{1/2}$, $8p_{1/2}$, and $9p_{1/2}$
states. The value $R_{\rm max}=75 a_B$ is large enough for the
$9p_{1/2}$ to be physical state. The number of splines is chosen to
be sufficiently large to saturate the summation. It turns out that
saturation is achieved at approximately 80 B-splines (in \cite{Porsev}
the authors used 40 B-splines of a different type).

%Note that since we are interested in the ratio of the {\em tail} to {\em main}
%contributions, one can argue that this ratio may depend on the basis even if the
%final answer is the same. This might be true if the issue is not
%properly treated. To fix the {\em tail}-to-{\em main} ratio the basis
To compare the tail terms in different calculations, the basis sets  
must satisfy two conditions. First, the box size and number of splines
must be large enough for accurate approximation of the all atomic
states entering main term so that these states can be associated with
real physical states. Second, the basis must be complete. For all
basis sets which satisfy both these conditions the tail
does not depend on basis. We believe that both our basis and that
used in~\cite{Porsev} satisfy these conditions. 

We include two types of the correlations, the core polarization effect
and Brueckner type correlations~\cite{CPM}. The core polarization is
the effect of the change in the self-consistent Hartree-Fock potential
due to external field. In our case we have two types of
external fields, the electric dipole field of the external photon and
the weak interaction of atomic electrons with the nucleus. As we
mentioned above, the inclusion of the core polarization is reduced to a
redefinition of the interaction Hamiltonians (plus small
``electroweak'' corrections considered in \cite{Nov,DFG02,CPM}). 
It is done in the
framework of the random phase approximation (RPA).

The Brueckner type correlations describe the correlation interaction of
the external electron with the atomic core, which can be reduced to
redefinition of the single-electron orbitals, constructing the 
Brueckner orbitals~\cite{CPM}. For this purpose we calculate 
correlation potential $\hat \Sigma$~\cite{CPM,Sigma-all} and
construct linear combinations of B-splines which are eigenstates of
the $\hat H_0 + \hat \Sigma$ Hamiltonian.

\section{Results and discussion}

%\begin{table}
%\caption{Contributions to the $E_{\rm PNC}$ [in $10^{-11}i(-Q_W/N)$
%    a.u.] for Cs in different approximations.}
%\label{t:pnc}
%\begin{ruledtabular}
%\begin{tabular}{lccr}
% & \multicolumn{2}{c}{This work} &\multicolumn{1}{c}{Ref.\cite{Porsev}} \\ 
% & \multicolumn{1}{c}{RPA\tablenotemark[1]} 
% & \multicolumn{1}{c}{BO\tablenotemark[2]} 
% & \multicolumn{1}{c}{CC\tablenotemark[3]} \\ 
%\hline
%Core ($n<6$)   & 0.0026 & 0.0018 & -0.0020 \\
%Main ($n=6-9$) & 0.8705 & 0.8711 &  0.8823 \\
%Tail ($n>9$)   & 0.0192 & 0.0238 &  0.0195 \\
%Total          & 0.8923 & 0.8967 &  0.8998 \\
%\end{tabular}
%\tablenotetext[1]{Core polarization but no correlations beyond it.}
%\tablenotetext[2]{Brueckner orbitals and core polarization.}
%\tablenotetext[3]{Coupled-cluster for the main term.}
%\end{ruledtabular}
%\end{table}

\begin{table}
\caption{Partial contributions to the $E_{\rm PNC}$ [in $10^{-11}i(-Q_W/N)$ 
    a.u.] for Cs in different approximations.}
\label{t:tail}
\label{t:pnc}
\begin{ruledtabular}
\begin{tabular}{lcccc}
\multicolumn{1}{c}{Approximation} & 
\multicolumn{1}{c}{Core} & 
\multicolumn{1}{c}{Main} & 
\multicolumn{1}{c}{Tail} & 
%\multicolumn{1}{c}{Tail/Main} & 
\multicolumn{1}{c}{Total} \\ 
\hline
RPA\tablenotemark[1]    &  0.0026 & 0.8705 & 0.0192 %& 0.0221 
& 0.8923 \\
BO($\hat \Sigma^{(2)}$)\tablenotemark[2]        &  0.0014 & 0.8612 & 0.0273
& 0.8897 \\ 
BO($\lambda\hat\Sigma^{(2)}$)\tablenotemark[3] &  0.0018 & 0.8709 & 0.0244
& 0.8971 \\ 
BO($\hat\Sigma^{(\infty)}$)\tablenotemark[4]        &  0.0018 & 0.8711 & 0.0238
& 0.8967 \\ 
BO($\lambda\hat\Sigma^{(\infty)}$)\tablenotemark[5] &  0.0018 & 0.8678 & 0.0242
& 0.8938 \\ 
Ref.~\cite{Porsev}\tablenotemark[6] & -0.0020 & 0.8823 & 0.0195 %& 0.0221 
& 0.8998 \\
\end{tabular}
\tablenotetext[1]{Core polarization but no correlations beyond it.}
\tablenotetext[2]{Brueckner orbitals (BO) calculated with the
  second-order $\hat\Sigma$.}
\tablenotetext[3]{BO calculated with rescaled
  second-order $\hat\Sigma$.}
\tablenotetext[4]{BO calculated with the
  all-order $\hat\Sigma$.}
\tablenotetext[5]{BO calculated with rescaled
  all-order $\hat\Sigma$.}
\tablenotetext[6]{DHF for the core term; coupled-cluster for the main term.}
\end{ruledtabular}
\end{table}

Table \ref{t:pnc} shows contributions to the PNC amplitude in Cs
obtained in different approximations within the sum-over-states
approach. Partial contributions will be considered in detail below.
Here we start our discussion with some general remarks.
The total PNC amplitude in the RPA approximation
obtained with the direct summation (DS) is
\[  E_{\rm PNC}^{\rm RPA}({\rm DS}) = 0.89235 \times 10^{-11}i(-Q_W/N) . \]
The total PNC amplitude in the this approximation obtained via solving
equations (SE) (\ref{eq:edpsi}) and then using formula (\ref{eq:pnc2}) is
\[  E_{\rm PNC}^{\rm RPA}({\rm SE}) = 0.89234 \times 10^{-11}i(-Q_W/N) . \]  
Note the excellent agreement between these two numbers. Since both methods of
calculation share very little in common, it is safe to assume that
numerical error is negligible in both cases and the number truly
represents the PNC amplitude in the RPA approximation.
This implies high quality of the basis used in present work and its
suitability to study partial contributions.

Table \ref{t:pnc} shows significant effect of correlations on the tail
contribution and significant difference in the values of core and tail
contributions between present work and \cite{Porsev}. There is also
1.3\% difference for the main term. However, this difference is not
important. Accurate treatment of the main term goes beyond the scope
of the present paper; we just take its value from \cite{Porsev}. If core
and tail terms also have this 1.3\% relative uncertainty then it would
be more than satisfactory for the purpose of the present work. We will
see below that actual uncertainty is higher.

\subsection{Core contribution}

\begin{table}
\caption{Contributions of the core states $np_{1/2}$ to the $E_{\rm
    PNC}$ [in $10^{-11}i(-Q_W/N)$ 
    a.u.] for Cs in different approximations.}
\label{t:core}
\begin{ruledtabular}
\begin{tabular}{lddd}
 & \multicolumn{1}{c}{$\langle \delta \psi_{6s} | {\bm d}| \psi_{7s} \rangle$}
 & \multicolumn{1}{c}{$\langle \psi_{6s} | {\bm d}| \delta \psi_{7s} \rangle$}
 & \multicolumn{1}{c}{Sum} \\ 
\hline
DHF                     & 0.02471  &  -0.02645  &  -0.00174 \\
RPA\tablenotemark[1]    & 0.05991  &  -0.05821  &   0.00170 \\
RPA\tablenotemark[2]    & 0.06043  &  -0.05784  &   0.00259 \\
BO+RPA\tablenotemark[2] & 0.07231  &  -0.07049  &   0.00182 \\
\end{tabular}
\tablenotetext[1]{RPA equations solved at Hartree-Fock frequency.}
\tablenotetext[2]{RPA equations solved at experimental frequency, $\omega=0.0844$ a.u.}
\end{ruledtabular}
\end{table}

%The difference in the results for the core contribution is an
%obvious source for the difference in final results of the present work
%and \cite{Porsev} and, in our view, the main reason for underestimation of
%the error bars in \cite{Porsev}.

Calculations in the present work are done in two different ways. One uses
the sum-over-states approach and limits the summation in
(\ref{eq:pnc}) to core states. Another uses the weak corrections to
the $6s$ and $7s$ states found by solving differential equations (see
formula (\ref{eq:pnc2})). The contribution of the core states are
found by imposing orthogonality conditions for $\delta\psi_{6s}$ and
$\delta\psi_{7s}$ and the states in the core. Both methods give
exactly the same results. 

To test the calculations even further we ran our code for the PNC in
Ra$^+$ and found excellent agreement with \cite{Ra+} for the core
contribution in RPA approximation.

Our results  are presented in Table \ref{t:core}. 
The result in the Dirac-Hartree-Fock (DHF)
approximation is in good agreement with those of \cite{Porsev} and
\cite{ND} (see also Table \ref{t:pnc}) which were calculated in the DHF
approximation in both works~\cite{Derevianko}.

Note the strong cancelation between terms with $\delta\psi_{6s}$ and
$\delta\psi_{7s}$. This cancelation makes the core contribution 
very sensitive to the inclusion of the core
polarization effect. We include it by solving the RPA equations for
both operators ($H_{\rm PNC}$ and ${\bm d}$). The equations for the
electric dipole operator are solved at $\omega = 0.0844$ a.u. which is the
experimental energy 
difference between the $6s$ and $7s$ states of Cs. 
 
The last line of Table~\ref{t:core}
presents the effect of using Brueckner orbitals for the core
contribution. The use of the Brueckner orbitals in the core can be
justified by the condition that core and valence
states must be orthogonal to each other and using the same $\hat
\Sigma$ operator in both cases is a good way to achieve this. The
difference in the core contribution using RPA and Brueckner
orbitals is relatively small (see Table~\ref{t:pnc}). We use this
difference as an estimate of the uncertainty of the core
contribution. 

The final difference between the present work and \cite{Porsev} for the
core contribution is 0.0038 in units of Table~\ref{t:core}. This
difference is mostly due to the core polarization effect. 
%The value of
%the difference is roughly 1.6 times the total uncertainty
%reported in~\cite{Porsev}.

\subsection{Tail contribution}

The third row of Table~\ref{t:tail} shows partial contributions
 to the tail component of the
PNC amplitude calculated in different approximations.
To include correlations we use four different sets of Brueckner
orbitals obtained with the use of two different correlation-correction
operators $\hat \Sigma$: the second-order operator $\hat \Sigma^{(2)}$
~\cite{CPM}; and the all-order operator $\hat \Sigma^{(\infty)}$
~\cite{Sigma-all}. Rescaling of $\hat \Sigma$ is done to fit the
energies of the lowest valence states. Rescaling usually improves the
wave functions and therefore the matrix elements.

%Inclusion of the correlations leads to significant increase in the
%values of the tail contribution and tail-to-main ratio (see Table
%\ref{t:tail}). Averaging over four last lines in the table gives
%0.0249(24) for the tail and  0.0287(28) for the tail-to-main ratio.
%Using this ratio and the result of \cite{Porsev} for the main term
%leads to the following value of 0.0253(25) for the tail contribution
%which is excellent agreement with our calculations.  

Inclusion of the correlations leads to significant increase in the
values of the tail contribution (see Table~\ref{t:tail}). 
For our final number we take our most complete calculation,
using the all-order $\hat\Sigma$ operator, while
the spread of values in the various Brueckner approximations gives a reliable
estimate of the uncertainty in our methods.

The result of \cite{Porsev} for the tail (0.0195, see Table
\ref{t:pnc}) was obtained using a blend of many-body
approximations including a simplified coupled-cluster method which
only includes single and double excitations \cite{Derevianko}. The
result is close to our result in the RPA approximation but
significantly smaller than our correlated value.
% It is interesting to
%note that the tail-to-main ratio in the RPA approximation is also very
%close to the result of \cite{Porsev}.

\subsection{Summary}

Table \ref{t:other} presents the results of the most accurate calculations
of the correlated (without Breit, quantum electrodynamic (QED) and
neutron skin correction) PNC 
amplitude in Cs. The abbreviation CP+PTSCI stands for correlation
potential \cite{CPM} combined with the perturbation theory in screened
Coulomb interaction, CC SD stands for coupled cluster with single and
double excitations, CC SDvT means coupled cluster with single, double
and valence triple excitations. All numbers, apart from those of
Ref.~\cite{Porsev} are in very good agreement with each other. But if
the result of \cite{Porsev} is corrected as shown in Table
\ref{t:pnc}, it comes to very good agreement with other calculations
as well (last line of Table~\ref{t:other}). 

\begin{table}
\caption{Correlated PNC amplitude in Cs [in $10^{-11}i(-Q_W/N)$
    a.u.] in different calculations. Breit, QED and neutron skin
    corrections are not included.}
\label{t:other}
\begin{ruledtabular}
\begin{tabular}{dl}
\multicolumn{1}{c}{Value} & 
\multicolumn{1}{c}{Source and method} \\ 
\hline
0.908(9)  & CP+PTSCI, Ref.~\cite{Nov} \\
0.909(9)  & CC SD, Ref.~\cite{ND} \\
0.905(9)  & MBPT with fitting, Ref.\cite{KPT01} \\
0.9078(45) & CP+PTSCI, Ref.~\cite{DFG02} \\
0.8998(24) & CC SDvT, Ref.~\cite{Porsev} \\
0.9079(40) & This work \\
\end{tabular}
\end{ruledtabular}
\end{table}

\begin{table}
\caption{All significant contributions to the $E_{\rm PNC}$ [in
  $10^{-11}i(-Q_W/N)$ a.u.] for Cs.}
\label{t:all}
\begin{ruledtabular}
\begin{tabular}{ldl}
\multicolumn{1}{c}{Contribution} & 
\multicolumn{1}{c}{Value} & 
\multicolumn{1}{c}{Source} \\
\hline
Core ($n<6$)   &  0.0018\,(8)  & This work \\
Main ($n=6-9$) &  0.8823\,(17) & Ref.~\cite{Porsev} \\
Tail ($n>9$)   &  0.0238\,(35) & This work \\
Subtotal       &  0.9079\,(40) & This work \\
%Subtotal       &  0.9078\,(45) & Ref.~\cite{DFG02} \\
Breit          & -0.0055\,(1) & Ref.~\cite{DeBreit,DzBreit} \\
QED            & -0.0029\,(3) & Ref.~\cite{QED} \\
Neutron skin   & -0.0018\,(5) & Ref.~\cite{DeBreit} \\
%$e-e$ weak interaction & 0.0003 & Ref.~\cite{
Total          & 0.8977\,(40) & This work \\
\end{tabular}
\end{ruledtabular}
\end{table}

We summarize the results in Table \ref{t:all}. We take the main term
from Ref.~\cite{Porsev} assuming that its value and uncertainty were
calculated correctly. The core and tail contributions come from the
present work.
% Note excellent agreement in the resulting subtotal amplitude
%with our previous calculations~\cite{DFG02}. 
Then we add all other
significant contributions to the PNC amplitude in cesium which can be
found in the literature. The final value for the PNC amplitude is
\begin{equation}
 E_{\rm PNC} = 0.8977\,(40) \times 10^{-11}i(-Q_W/N)\,,
\label{eq:final}
\end{equation}
which is in excellent agreement with our previous calculations, 
$E_{\rm PNC} = 0.8980\,(45)$~\cite{DFG02,QED}.
%If we take the correlated PNC amplitude from our previous
%calculations~\cite{DFG02} and all minor corrections (Breit, QED,
%neutron skin) as in Table~\ref{t:all} than the final PNC amplitude is
%\begin{equation}
% E_{\rm PNC} = 0.8976(45) \times 10^{-11}i(-Q_W/N).
%\label{eq:DFG}
%\end{equation}
The experimental value for the PNC amplitude is~\cite{Wood}
\begin{equation}
 E_{\rm PNC}/\beta = 1.5935\,(56) \ {\rm mV/cm}.
\label{eq:Wood}
\end{equation}
The most accurate value for the vector transition probability $\beta$
comes from the analysis~\cite{beta2} of the Bennett and Wieman
measurements~\cite{beta1} 
\begin{equation}
  \beta = 26.957\,(51)\, a_B^3.
\label{eq:beta}
\end{equation}
Comparing (\ref{eq:final}), (\ref{eq:Wood}) and (\ref{eq:beta}) leads
to
\begin{equation}
  Q_W(^{133}{\rm Cs}) = -72.58\,(29)_{\rm expt}\,(32)_{\rm theory}\,.
\label{eq:qw}
\end{equation}
This value is in a reasonable agreement with the prediction of
the standard model, $Q_W^{\rm SM} = -73.23\,(2)$~\cite{SM}. % being
%only about 1$\sigma$ smaller. 
If we add theoretical and experimental
errors in (\ref{eq:qw}) in quadrature, the Cs PNC result deviates from
the standard model value by 1.5\,$\sigma$:
\begin{equation}
 Q_W-Q_M^{\rm SM} \equiv \delta Q_W = 0.65\,(43).
\label{eq:dq}
\end{equation}
For small deviations from the Standard Model values we may relate this to the
deviation in $\sin^2\theta_W$ using the simple relationship
$\delta Q_W \approx -4Z\, \delta(\sin^2\theta_W)$ which gives
\begin{equation}
  \sin^2\theta_W = 0.2356\,(20) \,.
\label{eq:sint}
\end{equation}
This is also 1.5$\,\sigma$ off the standard model value
$0.2386\,(1)$~\cite{SM} at near zero momentum transfer.

The new physics originated through vacuum polarization to gauge boson
propagators is described by weak isospin conserving $S$ and isospin
breaking $T$ parameters~\cite{Rosner}
\begin{equation}
 Q_W-Q_M^{\rm SM} = -0.800\,S-0.007\,T.
\label{eq:ST}
\end{equation}
At the 1$\sigma$ level (\ref{eq:dq}) leads to $S=-0.81\,(54)$.

Finally, a positive $\Delta Q_W$ could also be indicative of an extra
$Z$ boson in the weak interaction~\cite{Marciano}
\begin{equation}
 Q_W-Q_M^{\rm SM} \approx 0.4(2N+Z)(M_W/M_Z{_{\chi}})^2.
\label{eq:Zxi}
\end{equation}
Using (\ref{eq:dq}) leads to $M_{Z_{\chi}}>710$ GeV/c$^2$.

\acknowledgments

The authors are grateful to A. Derevianko for useful discussions.
The work was supported in part by the Australian Research Council.

\end{document}